\documentclass[epj]{svjour}
\usepackage{graphics}
\usepackage{graphicx}
\usepackage{amsmath}
\usepackage{amsfonts}
\usepackage{amssymb}
\usepackage{hyphenat}
\usepackage[dvips]{color}
\usepackage{multicol}
\begin{document}
\title{Gravitational effects of free-falling quantum vacuum}
\author{Fran\c{c}ois~Couchot\inst{1},
Arache~Djannati-Ata\"{i}\inst{2},
Scott~Robertson\inst{3,1},
Xavier~Sarazin\inst{1},
Marcel~Urban\inst{1}%
}                     % Do not remove
\offprints{francois.couchot@ijclab.in2p3.fr}          % Insert a name or remove this line
\institute{Universit\'e Paris-Saclay, CNRS/IN2P3, IJCLab, 91405 Orsay, France
\and Universit\'e Paris Diderot, CNRS/IN2P3, APC, Paris, France
\and Universit\'e Paris-Saclay, Institut d'Optique Graduate School, CNRS, Laboratoire Charles Fabry, 91127 Palaiseau, France}
\date{Received: date / Revised version: date}
\abstract{
We present a model that builds ``dark matter"-like halo density profiles from free-falling zero-point vacuum fluctuations.
It does not require a modification of Newton's laws, nor the existence of as-yet-undiscovered dark matter particles.
The 3D halos predicted by our model are fully constrained by the baryonic mass distribution, and are generally far from spherical. 
The model introduces a new fundamental constant of vacuum, $T$, having the dimensions of time. 
We deduce the associated formalism from some basic assumptions,
and adjust the model successfully on several spiral galaxy rotation data while comparing our results to the existing analyses.
We believe our approach opens up a new paradigm that is worth further exploration, and that would benefit from checks relating to other phenomena attributed to dark matter at all time and distance scales.
Following such a program would allow the present model to evolve, and if successful it would make vacuum fluctuations responsible for the typical manifestations of dark matter.
%\PACS{
%     {PACS-key}{discribing text of that key}   \and
%      {PACS-key}{discribing text of that key}
%     } % end of PACS codes
} %end of abstract
\authorrunning{Fran\c{c}ois~Couchot {\it et al.}}
\titlerunning{Gravitational effects of free falling quantum vacuum}
\maketitle
\section{Introduction}
\label{intro}

Dark components of the Universe -- first dark matter, followed by dark energy -- arrived unexpectedly on the stage of modern physics a few decades ago.  Their reality seems undeniable, but there is still a great deal of speculation to be made concerning their origins.

As far as dark matter is concerned,
a first approach is to postulate the existence of unknown exotic particles from beyond the standard model, which interact with ordinary matter only gravitationally. 
Continuing research into these hypothesised dark matter particles has brought tremendous instrumental progress, but as yet no observational findings. 

A second approach to dark matter is to modify the effect of gravitation. 
For instance, the MOND paradigm~\cite{Milgrom-1983,Famaey-2012}, built as a 
dark matter-free explanation of the large distance 
flatness of the rotation curves of spiral galaxies, modifies Newtonian
dynamics in the low-acceleration regime by introducing a nonlinear term in the response of the acceleration to an applied force.
Approaches to found such an hypothesis on a theoretical basis
have been proposed, using a Lagrangian formalism~\cite{Skordis-2021}, and by interpreting it as an effect of the polarizability properties of a dark matter fluid~\cite{Blanchet-2007}.
On the other hand,
most of the modified gravitation models designed for dark energy have been excluded by an experimental measurement of the gravitational wave velocity, highly compatible with
$c$~\cite{Abbott-2017,Creminelli-2017}.

In this article, we propose a third conjecture that ascribes a definite gravitational role to the quantum
vacuum, and we show that this conjecture is compatible with the observed galactic rotation curves normally attributed to dark matter.

The vacuum plays a major role in many areas of modern physics.
Both of the towering monuments of $20^{\rm th}$ century theoretical physics give a key role to vacuum: General Relativity (GR) relates the geometrical properties of empty space to the stress-energy density distribution of all species~\cite{Misner-1973}, while Quantum Field Theory (QFT) relates the physical properties of vacuum to the whole set of elementary particles and symmetries of fundamental interactions~\cite{Bjorken-1965}. 
The photon component of the QED vacuum induces multiple effects:
the Casimir effect, where the photon vacuum is altered by the presence of conducting plates, leading to an attractive force between the plates;
the perturbative corrections to atomic energy levels ({\it e.g.}, the Lamb shift)
and to some particle properties ({\it e.g.}, the anomalous magnetic moment); 
and spontaneous emission, which can be interpreted as being sourced by the interaction of atomic excited states with the vacuum state of the electromagnetic field. 
The electron component of the QED vacuum is also predicted to play a role in, {\it e.g.}, inducing a macroscopic light-light interaction yielding an effective nonlinear optical index of vacuum, long sought-after by the PVLAS
experiment~\cite{PVLAS-2016} and more recently by the DeLLight project~\cite{Scott-2021}.
Renormalized QFT relies on vacuum to give finite values to
the effective masses and electrical charges of elementary particles, while fermion masses themselves are
considered as emerging from the interaction with the background Higgs field.
Regarding gravitation, quantum vacuum fluctuations are recognized to source black hole evaporation via Hawking radiation~\cite{Hawking-1974},
while the density inhomogeneities in today's Universe might have been seeded
by vacuum fluctuations parametrically amplified during the early stages of cosmic 
expansion~\cite{Mukhanov-2007}.

However, the proper treatment of vacuum fluctuations and their physical consequences remain poorly understood in certain contexts.
In particular, the influence of the vacuum energy density on the geometry of spacetime, calculated in the GR framework, is in contradiction with observations~\cite{Kragh-2012}: considering only its electromagnetic component, the uniform phase space density
integrated up to a Planckian cutoff yields a value higher than the measured density of our Universe by dozens of orders of magnitude. To escape this
paradox, one generally assumes that the vacuum energy 
is not a source of gravitation, which is not intellectually satisfying.
Only a few projects attempt to dig into the poorly explored gravitational properties of vacuum: one example is ARCHIMEDES~\cite{Calloni-2016}, which is based on the Casimir effect and aims to test whether the electromagnetic vacuum has a gravitational mass.

In this article, we present a new approach (to our knowledge) in which quantum vacuum is free-falling in gravitational fields, and fluctuating through a stationary stochastic process that 
allows the maintenance of stable vacuum density inhomogeneities wherever a gravitational field is present. We assume that these inhomogeneities are an extra source of gravitation, and show how they could replace dark matter halos on a galactic scale. This effect describes a ``dressing" of the Newtonian gravitational field, analogous to the Debye effect in plasmas where the medium adds extra forces to the standard electromagnetic interaction between bare charges~\cite{Saltzmann-1998}.

The paper is organised as follows.
In Section \ref{sec:model}, we present the assumptions of the model and derive the complete Poisson equation which takes the free-falling vacuum contribution into account.
In Section \ref{sec:SolGen}, we show how the model can be solved, and we apply it to the case of a point mass to show general properties of the solutions.
In Section \ref{sec:tests}, we test the model on the rotation curves of spiral galaxies, demonstrate the coherence of our results over a wide range of galaxy masses, and compare them to previous approaches.
The conclusion opens the way to future directions of research along similar lines.

%----------------------------------------------------------------------------------------------------------------------------------------------------------------------------------------------
\section{Model presentation
\label{sec:model}}
%----------------------------------------------------------------------------------------------------------------------------------------------------------------------------------------------
\subsection{Postulates
\label{sec:postul}}
%----------------------------------------------------------------------------------------------------------------------------------------------------------------------------------------------
We present here a set of initial assumptions, and propose to the reader that we derive the expected consequences were these assumptions to be realised.
 
Quantum vacuum is considered as a fluid, populated by elementary particles that are continuously popping in and out of existence. All space is assumed to be filled by this fluid, which is both sensitive to gravitation and a source of gravitation. We assume that, even if quantum vacuum is Lorentz invariant in a gravitation-free space,
its velocity field can be given a physical meaning in a gravitational context. We denote by
$\mathbf{v}(\mathbf{r})$ the flow velocity of this ``vacuum fluid", in a reference frame linked to a system of gravity source masses which are supposed stationary, {\it i.e.}, producing a time invariant gravitational field $\mathbf{g}^b$.
We denote such a reference frame as $\mathcal{R}_r$ (where the subscript $r$ stands for ``rest''). 
Although this vision of quantum vacuum is inferred from relativistic QFT, we treat the ``vacuum fluid" and its effects non-relativistically, restricting ourselves to $||\mathbf{v}||$ always much smaller than $c$. We thus stay in the weak-field regime of GR.

Inspired by the equivalence principle, we exploit the fact that, if the QFT vacuum state is to play a role in gravitation, its form should be easier to infer in Minkowskian free-falling reference frames. 
A special category of such frames, well-known in the analogue gravity
community, allows the description of the spacetime curvature in terms of a flowing fluid~\cite{Robertson-2012}. Around a point mass $M_0$, these frames 
move along radial trajectories of coordinate $r$, following a speed profile $\mathbf{v}_{\rm PG}$ given by
\begin{equation}
{\mathbf{v}_{\rm PG}^2(\mathbf{r})=-2\frac{G_N M_0}{r}=-2\, \Phi^b(\mathbf{r})\, ,}
\label{eq:V2phib}
\end{equation}
where $G_N$ is Newton's gravitational constant, and $\Phi^b$ the Newtonian potential around $M_0$. This velocity is the one which
enters the Painlev\'e-Gullstrand metric, giving an absolute meaning to $\mathbf{v}_{\rm PG}$ in a gravitational context.

We assume that the quantum vacuum is everywhere freely falling along such frames, generalized
to the total potential $\Phi$ which is simply the Newtonian
potential of a mass system, plus extra components emerging from the model.
We assume that this flow is everywhere directed towards decreasing potential. The absolute value of the flow speed is related to the total potential through
\begin{equation}
\boxed{\mathbf{v}^2(\mathbf{r})=-2\, \Phi(\mathbf{r})\,.}
\label{eq:V2U}
\end{equation}
This definition of $\mathbf{v}$ gives also an absolute meaning to the potential $\Phi$,
whereas potentials are usually defined up to an unphysical constant.
This is an important feature of the model.

Real $\mathbf{v}$ solutions of (\ref{eq:V2U}) exist only if $\Phi$ is negative.
We will restrict ourselves to such cases in this article.
Possible extensions of the model into
domains of positive potential will be addressed in a separate paper.

We assume that the free-falling quantum vacuum has a finite density, and
that this density is very slightly perturbed by the presence of gravitating
masses.  In the frame $\mathcal{R}_{\rm r}$, we express this assumption in the form
\begin{equation}
\rho(\mathbf{r})=\rho_0+\delta\rho(\mathbf{r})\, \ {\rm with} \ \ \delta\rho(\mathbf{r}) \ll \rho_0\, .
\label{eq:decomp}
\end{equation}
$\delta\rho(\mathbf{r})$ is the perturbation due to the action of the local gravitational field on the bulk density $\rho_0$.
Since $\rho_0$ is uniform, it is assumed {\it not} to create gravitational fields. We further assume that it has no measurable influence on the gravitational potential, whatever its (supposedly huge) value. Of course,
this is very different from GR, where even a constant energy density has a non-trivial effect on the metric. But we make this assumption in order to remain consistent with observations, which indicate that the large curvature expected does not in fact occur.

We assume that $\delta\rho$ originates from the fluctuating nature of
quantum vacuum placed in a gravitational field.
We postulate that the time correlation function of its ground state behaves in a non-trivial way, as does its electromagnetic component (which has recently been observed in a frequency range around one THz~\cite{Zurich-2019}).
We assume the existence of an effective fluctuation time, hereafter denoted
$\tau_0$. We expect it to be independent of $||\mathbf{v}||$.

\textbullet\ At time scales short compared to $\tau_0$, the vacuum fluid evolves like a classical medium, and obeys the classical continuity equation
\begin{equation}
\frac{\partial\rho}{\partial t}+\nabla\cdot\left(\rho\mathbf{v}\right)=0\, ,
\label{eq:Continuit}
\end{equation}
where the flow velocity $\mathbf{v}$ is constrained by (\ref{eq:V2U}).  
In general this yields $\nabla\cdot\mathbf{v}\ne 0$ in a non-uniform gravitational potential.
Expanding $\rho$ as in (\ref{eq:decomp}) to separate $\delta\rho$ from $\rho_0$ in (\ref{eq:Continuit}) leads
to the following equivalent equation for $\delta\rho$:
\begin{equation}
\frac{\partial\delta \rho}{\partial t}+\nabla(\delta \rho\mathbf{v}) =-
\rho_0\nabla\cdot\mathbf{v}\, .
\label{eq:develop3}
\end{equation}
This means that the effective fluid due to the perturbation $\delta\rho$ alone is not conserved on time scales
smaller than $\tau_0$.

\textbullet\ For durations greater than
$\tau_0$, one must account for the fact that the vacuum fluid regenerates, or relaxes, back to the constant density $\rho_0$.
This creates a source term on the right-hand side of (\ref{eq:Continuit}).
If this source term is equal to $\rho_0\nabla\cdot\mathbf{v}$, it cancels the right-hand term
of (\ref{eq:develop3}), which would mean that the effective fluid described by $\delta\rho$ is conserved on ``long" time scales, and can be considered as a standard gravitating medium.

The fluctuating nature of the quantum vacuum can be depicted from a semi-classical point of view: all space is filled with independent non-interacting virtual fermion-antifermion pairs
in a spin-0 state. This picture allows one to give a quantum origin to the electromagnetic properties of
vacuum~\cite{Leuchs-2013,Urban-2013}. This equivalent scalar field is looked at from a $\mathcal{R}_{\rm ff}$-type
Minkowskian frame, where pairs have an effective mean lifetime $\tau_0$ and an effective density $\rho_0$. During their lifetime, seen from $\mathcal{R}_{\rm r}$ they follow
on average the $\mathcal{R}_{\rm ff}$ trajectory and their density evolves along (\ref{eq:Continuit}), as for any fluid, which gives rise to a non-zero ${\delta\rho}$. But, when pairs
decay, the creation process regenerates new ones with a density reset to $\rho_0$, losing memory of the density inhomogeneities that had been induced earlier by now-decayed pairs.
The fluctuating quantum vacuum can be thought of as a continuous rain of evanescent drops, always recreated with the same density $\rho_0$, through a stationary stochastic process.

Following previous equations, we now derive the fundamental relation linking
the $\delta\rho$ and $\mathbf{v}$ fields. For time scales up to the characteristic fluctuation time $\tau_0$, 
(\ref{eq:develop3}) leads to
\begin{equation}
\frac{\partial\delta \rho}{\partial t}+\mathbf{v}\cdot\nabla\delta \rho =-
(\rho_0+\delta\rho)\nabla\cdot\mathbf{v}\, ,
\label{eq:developd}
\end{equation}
which on integration yields
\begin{equation}
{\delta\rho}(\mathbf{r}+\mathbf{v}\tau,t+\tau)-{\delta\rho}(\mathbf{r},t)=-(\rho_0+\delta \rho)\tau\,  \nabla\cdot\mathbf{v}\, .
\label{eq:deltarhorot}
\end{equation}
By assumption -- and as will be justified quantitatively below -- one can neglect $\delta\rho$ with respect to $\rho_0$ on the right-hand
side.

We intend to apply the model on astrophysical scales.
Since $\tau_0$ is a microscopic time scale, the falling distance
over a duration $\tau_0$ is tiny compared to the scale of spatial variation of $\delta\rho$.
This allows us to replace the material derivative with the time derivative,
leading to
\begin{equation}
{\delta\rho}(\mathbf{r},t+\tau)-{\delta\rho}(\mathbf{r},t)=-\rho_0\tau\,  \nabla\cdot\mathbf{v}\, .
\label{eq:deltarhorottoto}
\end{equation}
$\delta\rho$ builds up during the fluctuation timescale.
Starting from $\delta\rho=0$, and assuming that the time of relaxation is described by an exponential probability distribution with mean decay time $\tau_0$, we find
\begin{equation}
\left\langle \delta\rho \right\rangle = -\frac{\rho_0\, \nabla\cdot\mathbf{v}}{\tau_0}\int_0^\infty \tau\, e^{-\tau/\tau_0}  d\tau = -\rho_0\tau_0\,  \nabla\cdot\mathbf{v}\, .
\label{eq:deltarototo}
\end{equation}
Dropping the $\langle\rangle$ to simplify expressions, we find the microscopic relation between the 
vacuum density perturbation field
$\delta\rho$, and the free-falling velocity field $\mathbf{v}$:
\begin{equation}
\boxed{{\delta\rho}=-\rho_0\tau_0\,  \nabla\cdot\mathbf{v}\, .}
\label{eq:deltarhobar}
\end{equation}

The postulated mechanism produces density inhomogeneities proportional to the
divergence of the free-falling flow velocity.
In the presence of gravitational fields, these density inhomogeneities are not null. They can be
positive or negative, depending on the sign of the
divergence of $\mathbf{v}$.
Since $\mathbf{v}$ is proportional to the square root of the gravitational potential,
the density perturbation $\delta\rho$ produced by this mechanism
is not linearly linked to the potential: the solution corresponding to the sum of two different mass distributions is not the sum of the two distinct solutions.

%----------------------------------------------------------------------------------------------------------------------------------------------------------------------------------------------
\subsection{Complete Poisson equation
\label{sec:Poisson}}
%----------------------------------------------------------------------------------------------------------------------------------------------------------------------------------------------
We aim to apply our model in a weak-field regime and treat it classically, using
the Newtonian framework. 
The density perturbation ${\delta\rho}$, being a source of gravitation, enters
into the Poisson equation like a standard density, which leads to a ``complete" Poisson equation
that takes inhomogeneities of the free-falling vacuum into account.

In a frame $\mathcal{R}_{\rm r}$ linked to the source masses, we denote by $\Phi^b$ the Newtonian potential produced by a baryonic density $\rho^b$,
and the corresponding Newtonian acceleration field is denoted by $\mathbf{g}^b$ . These quantities are
linked by
\begin{equation}
\mathbf{g}^b=-\nabla\Phi^b\, ,
\label{eq:gradientb}
\end{equation}
and the Poisson equation
\begin{equation}
\Delta \Phi^b=-\nabla\cdot\mathbf{g}^b=4\pi G_N \rho^b\, .
\label{eq:Laplacien}
\end{equation}
In this article, we restrict ourselves to the simplest stationary case where, even if masses are moving, they
create a constant $\mathbf{g}^b$ field, as in the spiral galaxy case.

In the following, the source masses will be called ``free masses", in order to differentiate them from the extra masses drawn from the inhomogeneities of the falling vacuum.

We denote by $\delta \Phi$ the extra potential and by $\delta \mathbf{g}$ the extra acceleration field produced by ${\delta\rho}$.
Also, we write $\Phi \equiv \Phi^b+\delta \Phi$, $\rho \equiv \rho^b+{\delta \rho}$ and
$\mathbf{g} \equiv \mathbf{g}^b+\delta \mathbf{g}=-\nabla\Phi^b-\nabla\delta \Phi$.

The complete Poisson equation reads
\begin{equation}
\Delta \Phi=4\pi G_N \rho(\mathbf{r})\, .
\label{eq:Laplacetotal}
\end{equation}
Due to the linearity of the Poisson equation, ${\delta \rho}$ and $\delta \mathbf{g}$
are linked by the same relation as $\rho^b$ and $\mathbf{g}^b$, which yields
\begin{equation}
{\delta\rho}=- \frac{1}{4\pi G_N}\nabla\cdot\delta \mathbf{g}\, .
\label{eq:getrho}
\end{equation}
Using (\ref{eq:deltarhobar}) to replace ${\delta\rho}$ by its expression as a function of $\mathbf{v}$ gives a relation between $\delta \mathbf{g}$ and $\mathbf{v}$:
\begin{equation}
\rho_0\tau_0\,  \nabla\cdot\mathbf{v}= \frac{1}{4\pi G_N}\nabla\cdot\delta \mathbf{g}\, .
\label{eq:getrhoenv}
\end{equation}

Vector fields with the same divergence differ only by a divergence-free field.
A divergence-free $\delta \mathbf{g}$ component is considered as unphysical, since from (\ref{eq:getrho}) it would be produced by a null $\delta\rho$ field.
On the other hand, since a constant velocity field $\mathbf{u}$ has zero divergence, the model is invariant under the transformation $\mathbf{v} \to \mathbf{v}+\mathbf{u}$,
and the $\delta\rho$ solution is therefore unchanged if one gives a global
uniform motion to the full system (free masses plus vacuum). 

Setting $\mathbf{u}$ to zero (so that we effectively work in the centre of mass frame),
one gets a fundamental relationship between
$\delta \mathbf{g}$ and $\mathbf{v}$:
\begin{equation}
\delta \mathbf{g}=4\pi G_N\,\rho_0\tau_0\, \mathbf{v}\, ,
\label{eq:delgdev}
\end{equation}
where the microscopic model appears only through the product
$\rho_0\tau_0$. Introducing the characteristic time $T$:
\begin{equation}
T=\frac{1}{4\pi G_N\,\rho_0\tau_0}\, ,
\label{eq:definiT}
\end{equation}
leads to
\begin{equation}
\boxed{\delta \mathbf{g}=\frac{\mathbf{v}}{T}\,.}
\label{eq:delgdev2}
\end{equation}
$T$ is an effective time scale, governed by $\rho_0\tau_0$.  Noting that $c^{2} \rho_0 \tau_0$
has the dimensions of an action density, it may be associated to a ``natural" length scale
$\Lambda$, defined by the relation
\begin{equation}
\rho_0\tau_0=\frac{\hbar}{c^2\Lambda^3}\, .
\label{eq:density}
\end{equation}
Theoretical developments are needed to predict $T$ or $\Lambda$ from particle properties, but the lack of such a theory does not prevent us from testing the consequences of the model.
We show in Section \ref{sec:spiral} how they fit with spiral galaxy data, and extract estimates for $T$
in the range $\sim 10^{16}$ s, which gives order-of-magnitude estimates for $\Lambda$ and
$\rho_0\tau_0$:
\begin{equation}
\Lambda\approx 2\ {\rm fm, \ or\ }c^2 \rho_0\tau_0\approx 10^{-35}\ {\rm J.s/fm}^3\, .
\label{eq:ordermag}
\end{equation}
With a picture of fermion pairs filling the quantum vacuum, fluctuations are expected to
show up at time scales smaller than the lifetime associated to an $e^+e^-$ pair at rest, which is about
$\tau_e=\hbar/(4m_ec^2)\approx 3\times 10^{-22}$ s.
This leads to a lower bound for $\rho_0$:
\begin{eqnarray}
\rho_0  =  \frac{\rho_0\tau_0}{\tau_0} >  \frac{\rho_0\tau_0}{\tau_e}\ \approx\  . 2\ {\rm MeV/c}^2
{\rm /fm}^3\, .
\label{eq:bornerho}
\end{eqnarray}
Taking into account heavier fermion components would raise this lower bound on $\rho_0$.

The lower bound~(\ref{eq:bornerho}) is already sufficient to validate the assumption $\delta\rho\ll \rho_0$.
One can build  a natural density unit from the model parameters: $\rho_T=1/{G_N T^2}\approx 10^{-37}\ {\rm MeV/c}^2 {\rm /fm}^3$. This density shows up in the infinite plane solution, and the
range of $\delta\rho$ explored in this article lies within a few orders of magnitudes of $\rho_T$.
So $\delta\rho$ is, as announced, a tiny perturbation ($<10^{-36}$) of $\rho_0$.

Leaving the microscopic domain for the astrophysical one, we express mass densities in 
${M}_\odot/{\rm pc}^3$, using the equivalence $1\, {\rm MeV/c}^2{\rm /fm}^3 \equiv 3\times 10^{34}\, {M}_\odot
/{\rm pc}^3$. In these units, we have
\begin{eqnarray}
{\rho_T}  =  \frac{1}{G_N T^2}\approx 3\times 10^{-3}\ {M}_\odot/{\rm pc}^3\, .
\label{eq:rote}
\end{eqnarray}

%----------------------------------------------------------------------------------------------------------------------------------------------------------------------------------------------
\section{Solving the model}
\label{sec:SolGen}
%----------------------------------------------------------------------------------------------------------------------------------------------------------------------------------------------
\subsection{Initial elements}
We expand the total potential $\Phi(\mathbf{r})$ as the sum of three components:
\begin{equation}
\Phi(\mathbf{r})=\Phi^b(\mathbf{r})+\Phi_{\rm off}+\delta\Phi(\mathbf{r})
\label{eq:gradV2U}
\end{equation}

$\Phi^b(\mathbf{r})$ is the potential due to the local free mass system under study. It is the usual solution of the standard Poisson equation that tends towards zero as the distance to the studied mass system goes to infinity.

$\Phi_{\rm off}$ is an effective uniform offset potential. It is linked to the absolute value of an offset speed ${v}_{\rm off}$, defined by
\begin{eqnarray}
{v}_{\rm off}^2&=&-2\, \Phi_{\rm off}\, (\Phi_{\rm off}\le0)\, .
\label{eq:Voff}
\end{eqnarray}

To determine the solution $\delta\Phi(\mathbf{r})$,
we expand (\ref{eq:V2U}):
\begin{eqnarray}
\mathbf{v}^2  =  2 \left|\Phi^b({r})+\Phi_{\rm off}+\delta\Phi({r})\right|\, .
\label{eq:V2U3}
\end{eqnarray}
Since $\delta \mathbf{g}=-\nabla\delta\Phi = \mathbf{v}/T$,
we can express $||\delta \mathbf{g}||$
as a function of the potentials:
\begin{eqnarray}
||\delta \mathbf{g}|| & =&  \frac{1}{T} \sqrt{2\left|\Phi^b({r})+\Phi_{\rm off}+\delta\Phi({r})
\right|}\, .
\label{eq:dgdephi}
\end{eqnarray}
The vector $\mathbf{\delta g}$ points towards decreasing values of the potential.

We assume that, in the core of the mass distribution, the complete solution is close to the classical one. So, we impose the boundary condition
\begin{eqnarray}
\delta\Phi(\mathbf{r}_{\rm prox})=0\, ,
\label{eq:Boundary}
\end{eqnarray}
where $\mathbf{r}_{\rm prox}$ is the position of the nearest measured point to the
minimum of $\Phi^b$. 
This condition is not a strong one, since in individual adjustments, it can be
absorbed into a parameter linked to $\Phi_{\rm off}$.
It leads to the following self-consistent equation for $||\delta \mathbf{g}||$:
\begin{eqnarray}
||\delta \mathbf{g}|| & =&  \frac{1}{T} \sqrt{2\left|\Phi^b({r})+\Phi_{\rm off}
-\int_{\mathbf{r}_{\rm prox}}^\mathbf{r} \delta\mathbf{g}\cdot d\mathbf{s}
\right|}\, .
\label{eq:dgdephimin}
\end{eqnarray}

%**********************
\subsection{Spherically symmetric cases}
\label{sec:one}
\subsubsection
{$\delta g$ 1D differential equation}
\label{sec:diffeq}
In cases with spherical symmetry,
$\mathbf{v}$ is parallel to $\mathbf{g}^b$, both
being directed along the radial coordinate $r$.
The solution can then be obtained through the integration of a 1D differential equation.
Taking the derivative of equation (\ref{eq:V2U3}) with respect to $r$ leads to
\begin{equation}
\frac{d(v^2)}{dr}=  2\left(g^b+\delta g\right) \, ,
\label{eq:V2U4}
\end{equation}
We may eliminate $v$ using (\ref{eq:delgdev2}),
which leads to a differential equation for $\delta g$:
\begin{equation}
\frac{d(\delta g^2)}{dr}= \frac{2}{T^2}\left(g^b+\delta g\right) \,.
\label{eq:V2U5}
\end{equation}
This allows, along with the boundary condition (\ref{eq:Boundary}), a direct numerical integration.
We can also write it as
\begin{equation}
\frac{\partial (\delta g)}{\partial r}= \frac{1}{T^2}\left(1+\frac{g^b}{\delta g}\right)\, ,
\label{eq:equadifplat}
\end{equation}
which will give simple approximate solutions in some limiting cases.

In general, this equation must be solved numerically, except for the academic case of the infinite plane
where an analytical solution exists. 
This is useful as it allows us to validate the accuracy of the numerical code.

\subsubsection
{Point mass}
\label{sec:point}
Equation (\ref{eq:V2U5}) allows us to solve numerically the case of any
spherical mass distribution. In this section, we address the case of a point mass $M_0$
with baryonic potential $\Phi^b = -G_NM_0/r$. Even for this simplest example, analytical solutions exist only in
asymptotic cases. We summarize some properties of the solutions.

\paragraph{\it Analytical solution close to the point mass}\hfill
\vskip .2 cm
Equation (\ref{eq:equadifplat}) can be simplified at short distances, when
$\delta g\ll g^b$, becoming: 
\begin{equation}
\frac{d\delta g}{dr}\approx  \frac{1}{T^2}\left(\frac{g^b}{\delta g}\right) \, .
\label{eq:V2U7}
\end{equation}
Writing $g^b$ explicitly gives
\begin{equation}
\frac{d(\delta g)^2}{dr} = -\frac{2G_N M_0}{r^2T^2} \, .
\label{eq:V2U8}
\end{equation}
So, denoting an integration constant by $a$, we have
\begin{equation}
| \delta g | = \frac{\sqrt{2G_N M_0}}{T} \sqrt{\frac{1}{r}+a}\, ,
\label{eq:V2U9}
\end{equation}
Note that, since $\mathbf{v}$ is assumed to point in the direction of decreasing potential ({\it i.e.}, towards the point mass), (\ref{eq:delgdev2}) tells us that $\delta g$ is negative, like $g^b$. Denoting by $\delta v=\sqrt{r|\delta g|}$ the extra rotation speed component produced by this extra gravitational field, it is to be noticed that this solution leads to a value of $\delta v$ proportional to $M_0^{1/4}$. This is reminiscent of the Tully-Fisher relation~\cite{Tully-1977}. This might give a hint about a physical origin of this observed property of spiral galaxies, despite the fact that we are dealing here with a point mass instead of a flat circular mass distribution.

The integration constant $a$ can be found by neglecting $\delta\Phi$ in equation (\ref{eq:dgdephi}), close to $M_0$:
\begin{equation}
\delta g(r \to 0)\approx -\frac{\sqrt{2G_N M_0}}{T}\sqrt{\frac{1}{r}+\frac{v^2_{\rm off}}{2G_N M_0}}\, .
\label{eq:dg0}
\end{equation}
This shape is valid in a limited $r$ domain:
\begin{equation}
|g^b|=\frac{{GM_0}}{r^2}\gg \frac{1}{T}\sqrt{2(|\Phi_{\rm off}|+|\Phi^b|)}
\label{eq:limited}
\end{equation}

Equation (\ref{eq:getrho}) allows us to compute the extra density close to a point mass $M_0$:
\begin{equation}
\lim_{r \to 0} \delta \rho = \frac{3}{4\pi G_N T}\sqrt{\frac{G_N M_0}{2r^3}}
\label{eq:rho0dereq0}
\end{equation}
which is positive, and varies as $r^{-3/2}$. It creates an extra mass
proportional to $r^{3/2}$ when integrated from $0$ to $r$.
Although this halo shape around a point mass is of the cusp type, like those of the $\Lambda$CDM 
model~\cite{deBlok-2009}, we will see in Section \ref{sec:Gaz} that, for complex mass distributions, our
model reproduces well some rotation curves best interpreted up to now as being produced by a core-type
halo. 

\paragraph{\it Numerical solution}\hfill
\vskip .2 cm
\begin{figure}
\centering
\resizebox{0.48\textwidth}{!}{%
  \includegraphics{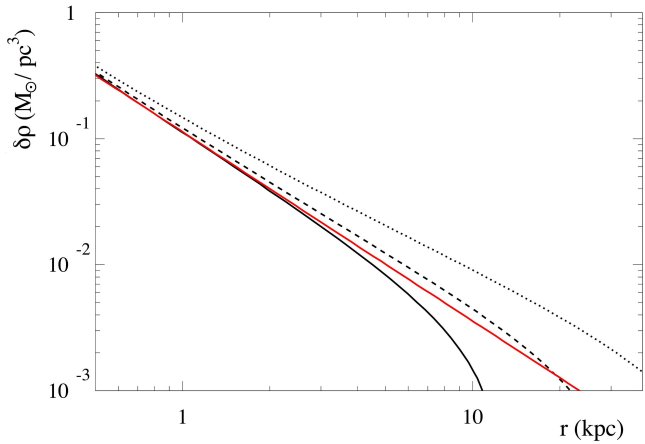}
}
\caption {\label{fig:dens}{Numerical solution of $\delta \rho(r)$ around a point 
mass $M_0 = 10^{11}\ M_\odot$, with $T=7\times 10^{15}\,$s, and for $v_{\rm off}=0$ (black
continuous curve), $300$ km/s (dashed curve) and $600$ km/s (dotted curve). The red line is the
approximate solution (\ref{eq:rho0dereq0}).}}
\end{figure}
Figure \ref{fig:dens} shows the 
solution $\delta\rho(r)$ around a point mass of $10^{11}$ solar masses, with
$T=7\times 10^{15}\,$s, for several values of $v_{\rm off}$. The red curve shows
the asymptotic solution  (\ref{eq:rho0dereq0}), well in agreement with the solution for $v_{\rm off}=0$ at small $r$ values. $\delta\rho$ depends on $v_{\rm off}$, being
stronger in a deeper potential offset.
\begin{figure}
\centering
\resizebox{0.48\textwidth}{!}{%
\includegraphics{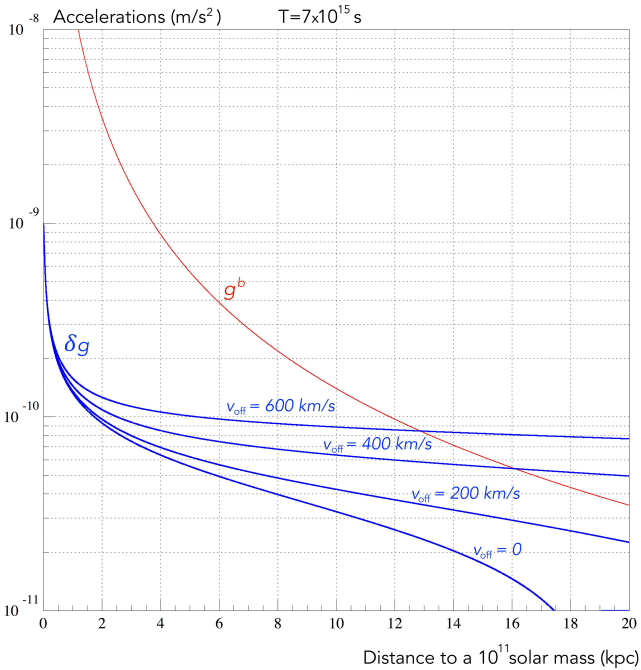}
}
\caption {Examples of spherical solutions: in red the Newtonian acceleration $g^b(r)$, in blue $\delta g(r)$ for different values of $v_{\rm off}$, with $T=7\times 10^{15}\,$s.}
\label{fig:pointa}
\end{figure}
The absolute values of $g^b(r)$ and $\delta g(r)$, computed under the same conditions, are shown for several values of $v_{\rm off}$ on Figure \ref{fig:pointa}, and for several values of $T$ on Figure \ref{fig:pointb}.
In this $(T,v_{\rm off})$ range, the extra field can dominate over the Newtonian one for distances greater than about $10$ kpc.

\begin{figure}
\centering
\resizebox{0.43\textwidth}{!}{%
  \includegraphics{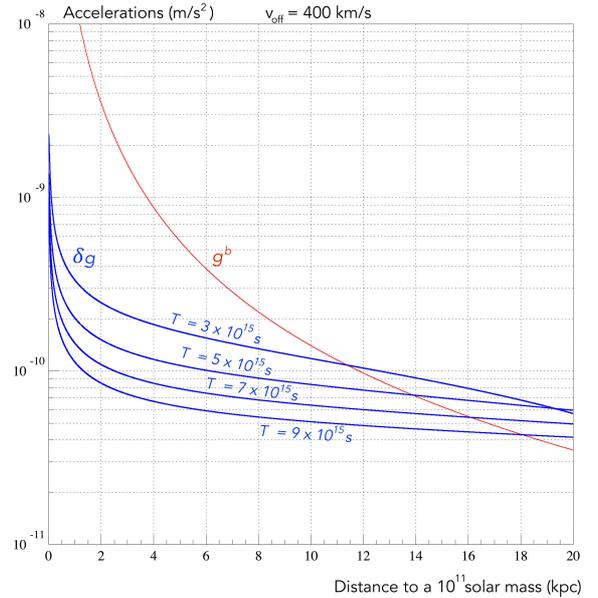}
}
\caption {\label{fig:pointb}{Examples of spherical solutions: in red the Newtonian acceleration $g^b(r)$, in blue $\delta g(r)$ for different values of $T$, with $v_{\rm off}=400$ km/s.}}
\end{figure}

\begin{figure}
\centering
\resizebox{0.43\textwidth}{!}{%
\includegraphics{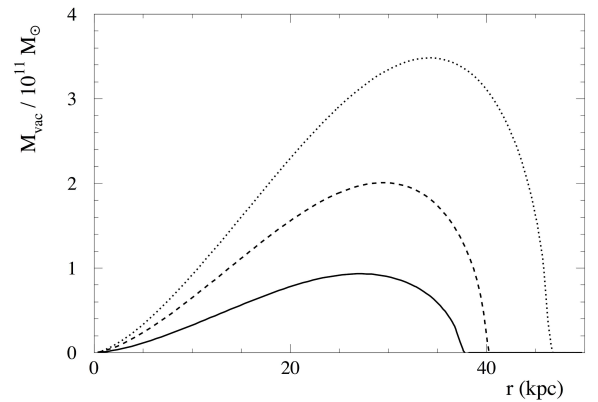}}
\caption {\label{fig:mvac}{Numerical solution of $M_{\rm vac}(r)$ for $M_0=10^{11}\ M_\odot$ (black
continuous curve), $5\times 10^{11}\ M_\odot$ (dashed curve) and $10^{12}\ M_\odot$  (dotted curve), with $v_{\rm off}=200$ km/s and $T=7\times 10^{15}\,$s.}}
\end{figure}

As expressed in~(\ref{eq:deltarhobar}), the overdensity $\delta\rho$ is proportional to the divergence of the flow velocity $\mathbf{v}$ of the free-falling vacuum. Under spherical symmetry, this divergence is the sum of two terms:
\begin{equation}
\nabla\cdot\mathbf{v}=\frac{\partial v}{\partial r}+2\, \frac{v}{r}\, .
\label{eq:divenr}
\end{equation}
For negative values of $v$, the second term is negative, thus giving a positive contribution to the overdensity. But the first term is positive, so depending on the
ratio between both terms the resulting $\delta\rho$ can have any sign. As we have seen,
it is positive close to $M_0$. It changes sign at a distance hereafter called $r_{-}$.

We define the effective mass $M_{\rm eff}$, as the mass that would create the field
$\mathbf{g}=\mathbf{g}^b+\delta \mathbf{g}$ at distance $r$. We have
\begin{eqnarray}\nonumber
M_{\rm eff}(r) & = & M_0+M_{\rm vac}(r)\\
{\rm with\ \ } M_{\rm vac}(r) & = & r^2\frac{\delta g(r)}{G_N}=
4\pi\int_0^r \xi^2\delta\rho(\xi)d\xi\, .\  \ \  
\label{eq:Meff}
\end{eqnarray}
$M_{\rm vac}$ is the extra mass produced by the falling vacuum.
Near $M_0$, $M_{\rm eff}$ is a growing function of $r$, up to $r_{-}$ where,
due to the change in sign of $\delta\rho$, it starts to decrease. Its shape depends on
$M_0$ and $v_{\rm off}$. Figure \ref{fig:mvac} shows $M_{\rm vac}(r)$ for three values of $M_0$ and  $v_{\rm off}=200 $ km/s.

The rotation speed $v_{\rm rot}(r)$ of a small mass object orbiting
circularly at a distance $r$ around $M_0$  is
linked to $M_{\rm eff}(r)$ by the usual relation:
\begin{equation}
v_{\rm rot}=\sqrt{\frac{G_N M_{\rm eff}}{r}}\, .
\label{eq:vrot}
\end{equation}

Since $\delta g$ is (like $g^b$) directed inwards, $\delta\Phi$ is a growing function of $r$.
The domain for which the total potential stays negative is limited to a sphere of radius $r_\emptyset$.
On the surface of this sphere, $\Phi=v=\delta g=0$, so $M_{\rm eff}=M_0$. The integral of
$\delta\rho$ up to $r=r_\emptyset$ is exactly zero.

\begin{figure}
\centering
\resizebox{0.43\textwidth}{!}{%
\includegraphics{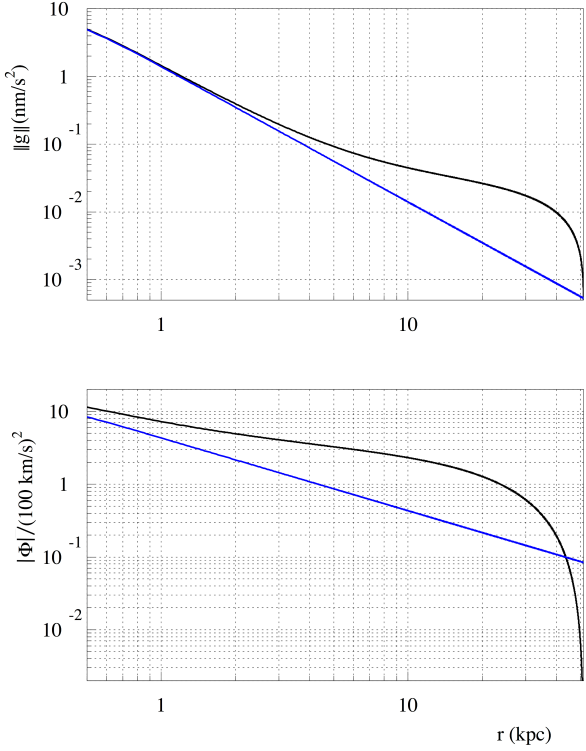}}
\caption {\label{fig:sph}{Numerical solutions of $||\mathbf{g}(r)||$ and $|\Phi(r)|$ around a point mass $M_0=10^{10}\,M_\odot$, with 
$T=7\times 10^{15}\,$s and $v_{\rm off}=250$ km/s. The purely baryonic acceleration and potential are shown in blue.}}
\vskip .2 cm
\end{figure}
\begin{figure}
\centering
\resizebox{0.43\textwidth}{!}{%
\includegraphics{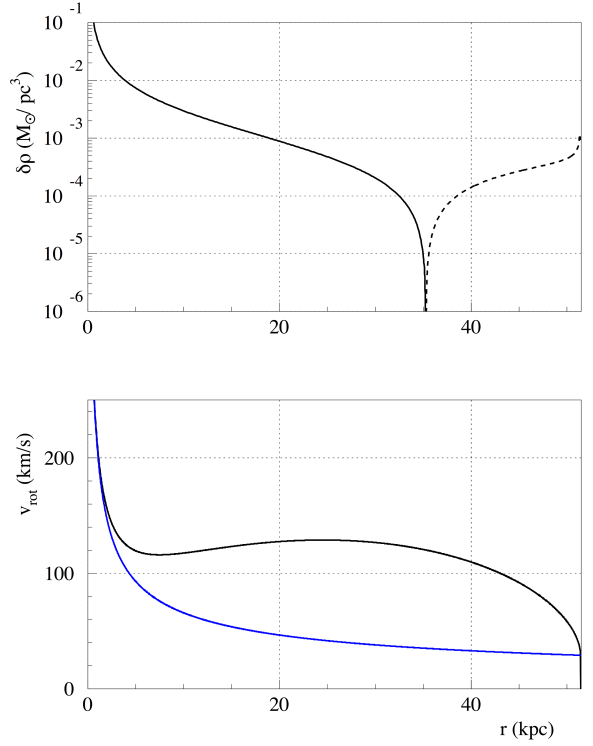}}
\caption {\label{fig:sph2}{Numerical solutions of $\delta\rho(r)$ and $v_{\rm rot}(r)$, for the same conditions as in Figure \ref{fig:sph}. Negative $\delta\rho$ values are shown by their absolute value as a dashed line. In this example, $r_{-}\approx 35$ kpc. The baryonic rotation speed is represented in blue.}}
\vskip .2 cm
\end{figure}

Figure \ref{fig:sph} shows $|g|$ and $|\Phi|$, and Figure \ref{fig:sph2} shows $\delta\rho$
and $v_{\rm rot}$, up to
$r_\emptyset$. The rotational velocity is larger than the Newtonian one. It plateaus over a relatively large region but does not reach an asymptotic value, which might lead to
predictions different from those of MOND at large distances.

\subsection{General case - Iterative solution}
\label{sec:two}
\subsubsection
{Reminder of basic equations}
\label{sec:recall}
In general, when the mass distribution has no particular symmetry, $\mathbf{v}$ and
$\mathbf{g}^b$ are not parallel.
$||\mathbf{v}||$ is still self consistently given by (\ref{eq:V2U3})
\begin{eqnarray}
\nonumber
\mathbf{v}^2  =  2 \left|\Phi^b({r})+\delta\Phi({r})+\Phi_{\rm off}\right|\, ,
\end{eqnarray}
recalling the fundamental relation between $\delta\Phi$ and $\delta \mathbf{g}$:
\begin{equation}
\nonumber
\delta \mathbf{g}=-\nabla\delta\Phi\,,
\label{eq:gradientdel}
\end{equation}
and that between $\delta \mathbf{g}$ and $\mathbf{v}$:
\begin{equation}
\nonumber
{\delta \mathbf{g}=\frac{\mathbf{v}}{T}\,.}
\label{eq:delgdev2bis}
\end{equation}
So, solving for $\delta \mathbf{g}$ is equivalent to solving for $\mathbf{v}$ or
for $\delta \Phi$. Given $\Phi^b(\mathbf{r})$, these equations have a unique solution
$\delta \Phi(\mathbf{r})$.

$\delta\rho,$ despite being the source of $\delta \mathbf{g}$, appears as a by-product
in the solving process for $\delta \mathbf{g}$. One can reformulate (\ref{eq:deltarhobar})
using $T$:
\begin{equation}
{\delta\rho}=- \frac{1}{4\pi G_NT}\,\nabla\cdot \mathbf{v}\, .
\label{eq:roger}
\end{equation}

\subsubsection
{Method}
\label{sec:methspace}
In the general case,  the $\mathbf{v}$ direction needs to be determined in each point,
since it is not aligned with the $\mathbf{g}^b$ direction. 

At first glance, it may seem that the problem can be expressed locally in the form of differential
equations, as in the effectively 1D case of spherical symmetry. We
searched for a local differential expression of the combined equations recalled above (see
Section \ref{sec:recall}). But, even making
use of the fact that $\delta\mathbf{g}$ is curl-free, the local system is underdetermined.
The full solution appears to depend on the full history of the free-falling frames, starting from places far away from the mass system under study. This intrinsic non-local character of the solution might be put in relation with the success met by non-local gravity models~\cite{Maggiore-2014} on all cosmological probes.

To determine the solution, we require a boundary condition. We surmise that $\mathbf{v}$ is parallel to $\mathbf{g}^b$
at a sufficiently large distance from the system of free masses.
One can then build up  the direction everywhere by letting frames fall into the full gravitational acceleration field
$\mathbf{g}=\mathbf{g}^b+ \delta\mathbf{g}$. Of course, $\delta\mathbf{g}$ being unknown,
this process has to be initiated and iterated until a stable $\delta\mathbf{g}$ solution is reached.

We solve for $\delta\Phi\, \&\, \delta\mathbf{g}$ iteratively, starting from $$\delta\Phi
\,\&\,\delta\mathbf{g}\equiv 0\,.$$
At each iteration, the free-falling velocity field is obtained by following fall trajectories from all directions,
as technically detailed below.
The $\delta\mathbf{g}$ field is derived directly from the $\mathbf{v}$ field using (\ref{eq:delgdev2}).
Then, integrating $\delta \mathbf{g}$ allows us to update
the $\delta\Phi$ solution.
The process proves to converge rather quickly ($\sim$10 iterations or so).

Technically, space is quantized on a mesh adapted to the problem
to be solved. $\mathbf{g}^b$ and $\Phi^b$ are first computed on all
mesh vertices. Free-falling trajectories start from a spherical surface of large radius, with an initial
speed given by (\ref{eq:V2U3}) and a direction aligned with $\mathbf{g}^b$ interpolated at every starting point from the closest mesh vertices. The angular sampling of trajectories is
thinner than the mesh sampling, and the time sampling allows to average a few points in every
mesh vertex.
In every sampled point, $\mathbf{g}^b$  is interpolated from the
closest mesh vertices. The $\mathbf{v}$ direction is determined on every vertex
as an average over all neighbouring samples weighted
according to their respective distances to the closest vertex. 

Every free-falling trajectory is stopped when it reaches a minimum of the potential.
This prescription is necessary for the model to reproduce observations. For now this is adopted only as
a recipe, as is the choice of the specific generalized Painlev\'e-Gullstrand free-falling reference
frames $\mathcal{R}_{\rm ff}$. A justification of these features would strengthen the model.
%One can notice that (\ref{eq:V2U}) gives a key role to the gravitational potential, acting on the
%quantum phase, as described for instance in a proposed gravitational Aharonov-Bohm experiment~\cite{Zeilinger-2012}.
%{\color{red} \it (SR: The last sentence is unclear.  What do you mean, exactly?)}

%----------------------------------------------------------------------------------------------------------------------------------------------------------------------------------------------
\section{Model test and adjustment on spiral galaxies}
\label{sec:tests}

Spiral galaxies are among the objects requiring the existence of dark halos, both for stability requirements, and for rotation curves (RC) consistency~\cite{Peebles-2017}.
Their study has been a very active field of research since the 70s, when the dark matter halo concept
appeared~\cite{Bertone-2016}.
Remarkable progress has been achieved both on the rotation curve measurements and on the galaxy baryonic mass modeling.

Their rotation curves have been interpreted via two different paradigms:
\begin{itemize}
\item{The dark matter halo paradigm: Extra mass emitting no light is present around the baryonic
galactic mass distribution, enhancing the gravitational field with respect to that due to baryons alone.  Making minimal
hypotheses on the nature of the dark matter, different halo profiles, mostly spheroidal, have been compared
to measurements and reproduce well the observations.}
\item{The MOND paradigm~\cite{Milgrom-1983}:
Newton's second law is modified in the
low-acceleration regime, applicable to stars in galactic outskirts (typically $a <10^{-10}$ m/s$^2$). This approach is successful on rotation curves, and it predicts the Tully-Fisher relation between luminous mass
and asymptotic speed, obeyed by spiral galaxies. It does not, {\it stricto sensu}, require
a galactic dark halo.
}
\end{itemize}
In the model we propose, halo density profiles originate from 
free-falling quantum vacuum fluctuations. This is achieved within the classical Newtonian framework, and within the standard model of known particles. The 3D halos
predicted by this model are fully constrained by the baryonic mass distributions,
and by the values of the offset potential and the parameter $T$. They are not spherical,
except for spherically distributed free masses.

%----------------------------------------------------------------------------------------------------------------------------------------------------------------------------------------------
\subsection{Framework
\label{sec:spiral}}
%----------------------------------------------------------------------------------------------------------------------------------------------------------------------------------------------

We adopt cylindrical coordinates:
the normal to the galactic plane is taken to be the $z$-axis, while $r$, the galactocentric radius,
is the distance from any point in the $z=0$ plane to the galactic centre.
For all studied galaxies, we use azimuthally symmetric baryonic matter distribution models. These galactic modelings give estimates of the $(r,z)$ mass distribution for a set of baryonic components
(usually one stellar and one gaseous disk, and one star bulge).
In most cases, we use the flat approximation. For M33, we used the thick description
of the galactic plane, with a flared disk, symmetrical with respect to the galactic plane. 

Due to the symmetries, in the galactic plane $z=0$ the problem becomes effectively 1D,
with $\mathbf{g}^b$ reducing to its radial component $g^b$,
and we can apply the method described in the spherical case (section \ref{sec:one}).

The solution $\delta g(r)$ is obtained through the integration of
(\ref{eq:V2U5}), starting from $r_{\rm prox}$, the smallest available galactocentric coordinate
in rotation curve data. One sets arbitrarily $\delta\Phi(r_{\rm prox})=0$.
Relation (\ref{eq:dgdephimin}) can be rewritten as:
\begin{eqnarray}
\delta g(r_{\rm prox}) & = &  -\frac{v_{\rm prox}}{T}\,\\
\nonumber{\rm \ with}\  v_{\rm prox}^2  & = & -2\Phi^b(r_{\rm prox})+v^2_{\rm off}\, .
\label{eq:vfree}
\end{eqnarray}
$v_{\rm prox}$ is the fit parameter that sets the boundary condition on  $\delta g(r_{\rm prox})$.
A complete galactic baryonic model can then allow to separate
$v_{\rm off}$ from $\Phi^b(r_{\rm prox})$. This is the case for M33, for which we derive 
$v_{\rm off}$ from the fit and the available baryonic model. For all other galaxies studied in this
article, we give only the $v_{\rm prox}$ value output by the fit.

For comparative purposes, we used the same modelings as in the published
dark halo or MOND analyses on the same galaxies.

This 1D method is much faster than the full 2D iterative solution described in Section \ref{sec:SolGen}. We checked on certain examples that both methods give the same results.

As in usual approaches of galactic halos, we assume that the star movement is purely rotational, and
neglect the radial velocity component. Such non circular movements are always present at some level. They can be the source of oscillations in the residuals of the fitted rotation curve and, in some cases, even 
require to rule out data points close to the galactic center.

Once integrated, the solution for $\delta g$ is converted into $\delta(v^2)$ via
\begin{equation}
\delta g=-\frac{\delta(v^2)}{r}\, .
\label{eq:gv2}
\end{equation}
This extra rotation speed is added quadratically to the Newtonian one, which allows us to compare galactic rotational data to those expected from the model, and thereby to extract estimates of the parameters $T$ and $v_{\rm off}$
based on a chi-square minimisation.

If gaseous galactic components are relatively well measured thanks to the characteristic lines in
spectra that give a direct access to column densities, we need to include unknown scaling factors on the stellar mass to light ratios, similarly to other analyses of the same objects.

The knowledge of galaxy evolution, through stellar po\hyp{}pulation synthesis models, allows us to reproduce the non-uniform stellar properties as a
function of the galactocentric radius, but an overall uncertainty on the M/L ratio remains for each component (bulge and star disk). We multiply the M/L ratio given by the models
by an adjustable factor denote $\Upsilon_c$.

We first present the fit on M33, a well-modeled and precisely measured galaxy.  We then apply the
method to two galaxy sets covering a large range in mass, and
already analyzed by the usual approaches.
\subsection{M33}
\label{sec:M33}
M33 is a nearby light spiral galaxy, with an estimated total baryonic mass around $10^{10}$
solar masses. This is one of the best measured spiral galaxies, both
concerning the rotation curve and the mass modeling. Its distance (840 kpc) is known
with a 4\% precision~\cite{Gieren-2013}, adding a negligible uncertainty in the conversion from
angular to linear distances.

Its rotation curve and mass model have been updated in 2014~\cite{Corbelli-2014} and used or improved in subsequent analyses~\cite{Lopez-2017,Banik-2020}.
Its baryonic model consists of a 2-component disk made of gas and stars.

We base our analysis on the RC and mass model of Ref.~\cite{Corbelli-2014} ($BVIgi$
stellar mass model, and thick disks).
\begin{figure}[htb]
\resizebox{0.48\textwidth}{!}{%
\includegraphics{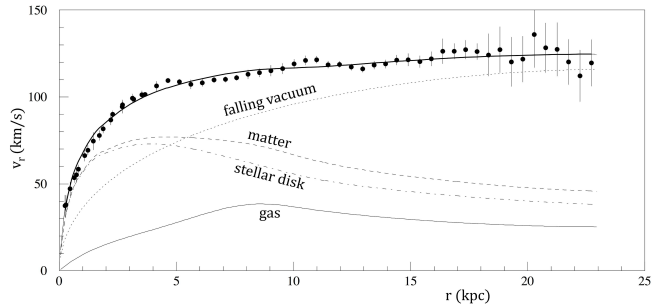}}
\caption {\label{fig:M33}{M33 rotation curve $v_r(r)$ (black dots with error bars) with the fit result
(continuous line), 
and the contributions from gas, stars, total matter, and ``falling vacuum'' labelled.}}
\end{figure}

\begin{figure}
\centering
\resizebox{0.48\textwidth}{!}{%
\includegraphics{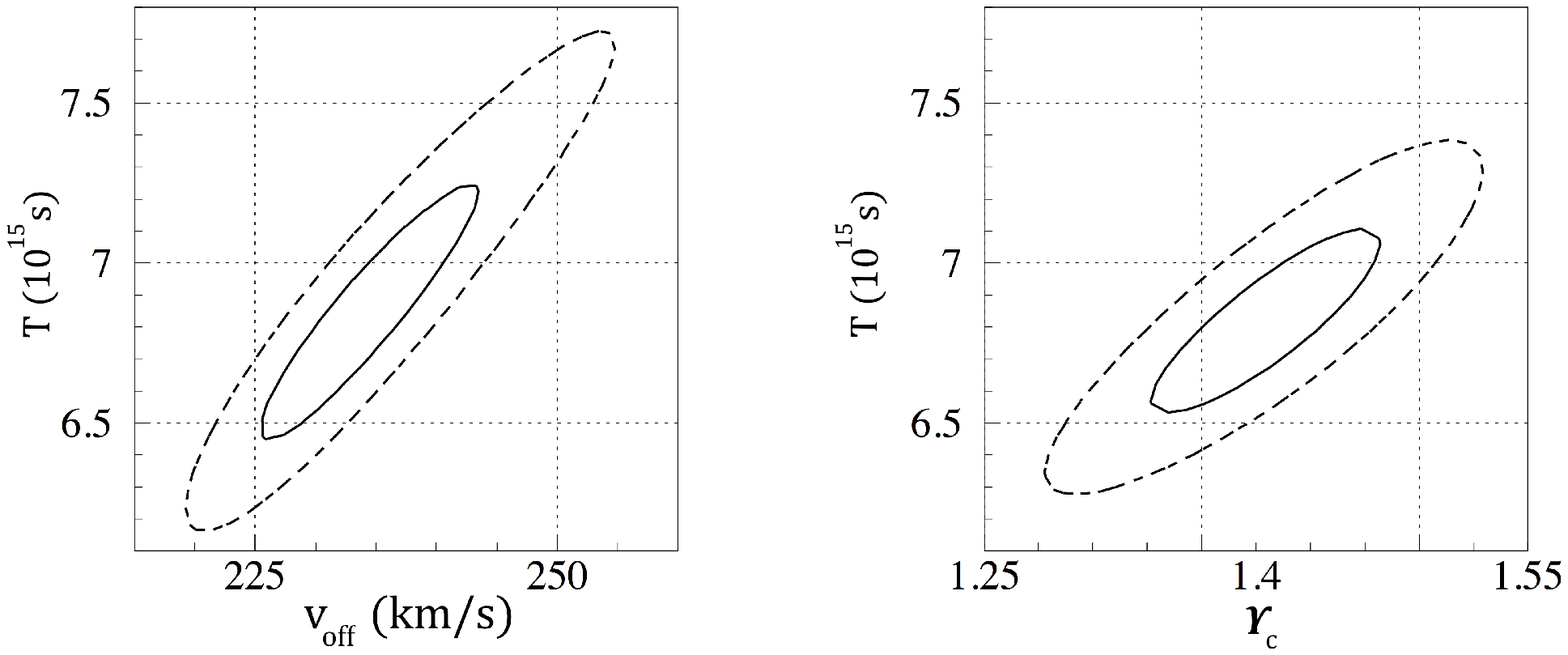}}
\resizebox{0.27\textwidth}{!}{%
\includegraphics{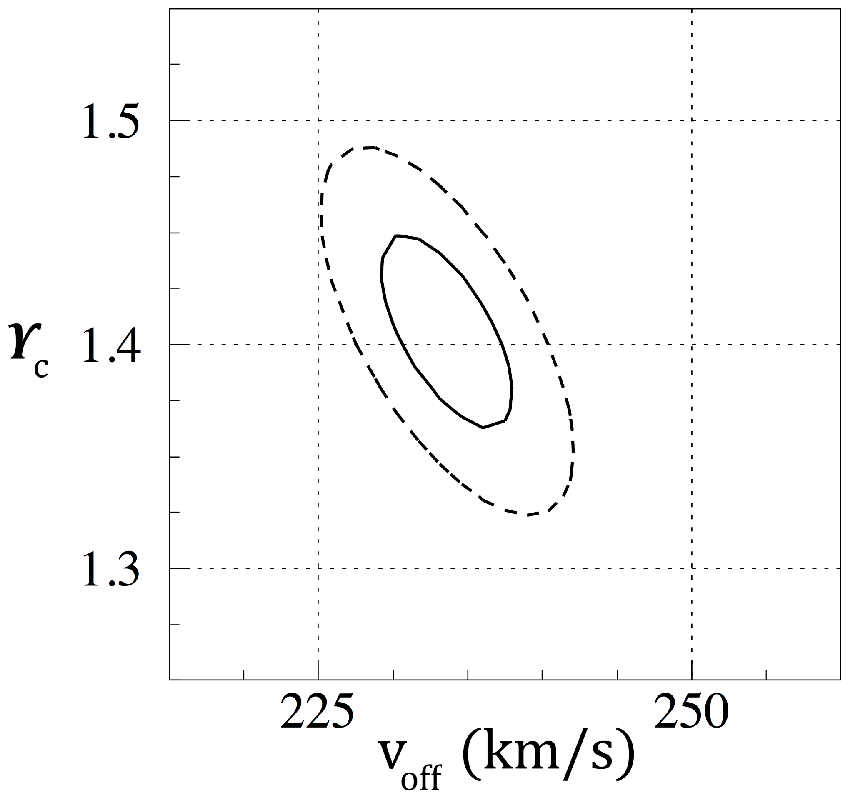}}
\caption {\label{fig:correl}{Correlations of the fitted parameters. Continuous (resp. dashed) line is
the 2D 68 $\%$ (resp. 95 $\%$) Confidence Level contour.}}
\end{figure}

The adjustment successfully matches the RC with no explicit dark matter halo, as shown in Figure \ref{fig:M33}. It returns the following estimates for $T$, $v_{\rm off}$ and $\Upsilon_c$:
\begin{equation}
  \left\{
      \begin{aligned}
\ \ T &= &  (6.8 &\pm.6)\times 10^{15}& &{\rm s}\ \ \ &\\
\ \ v_{\rm off} &=&    \ 230 &\pm 10& &{\rm km/s}\ \ \ &\\
\ \ \Upsilon_c &=&    \ 1.4 &\pm .06& &\ \ \ &\\
      \end{aligned}
    \right.
\label{eq:fitM33}
\end{equation}

$\Upsilon_c$ is compatible with the upper limit given by the stellar disk model~\cite{Corbelli-2003,Kam-2017} and gives a disk stellar mass out to $23$ kpc of $(6.8\pm .3)\times 10^{9}\ M_\odot$.

$v_{\rm off}$, like $v_r$, stays in the non-relativistic domain, as anticipated.
This is the case for all fitted galaxies in this article.

The $\chi^2$ is equal to 57 for 55 degrees of freedom, 
 giving a reduced chi-squared of $\chi_{\rm r}^{2} = 57/55 = 1.04$.

The 2D 68$\%$ and 95$\%$ Confidence Level contours for $T,\ v_{\rm off}$ and $\Upsilon_c$ are shown in Figure 
\ref{fig:correl}. $T$ is rather strongly correlated with the other parameters.
Once the 3 parameters are fixed to their best estimates by the 1D ($z=0$) fit,
we run an iterative 2D vacuum fall,
as described in Section \ref{sec:methspace}, which allows us to compute the full gravitational
and density fields away from the $z=0$ plane. The free-falling trajectories are begun with $\mathbf{g}^b$ and $\delta \mathbf{g}$
aligned at $32$ kpc from the galactic centre.

\begin{figure}
\centering
\resizebox{0.39\textwidth}{!}{%
\includegraphics{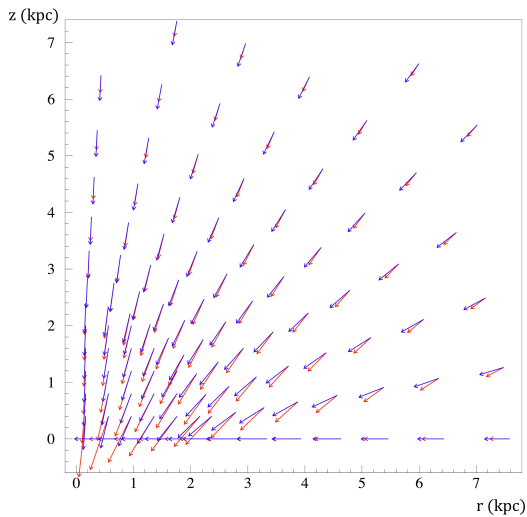}}
\caption {\label{fig:gs}{Cylindrical cut of $\mathbf{g}^b(r,z)$ (red arrows) and $\delta \mathbf{g}(r,z)$ (blue arrows) fields around the centre of M33.  The
galactocentric distance $r$ is the horizontal coordinate, and the distance $z$ to the galactic plane is the vertical coordinate. The arrow lengths are proportional to
the square roots of the field strengths.}}
\end{figure}

Figure \ref{fig:gs} shows $\mathbf{g}^b$ and $\delta \mathbf{g}$ in a small region around M33's centre.  The arrow lengths are proportional to
the square root of the fields, which allows the covering of the whole region with comparable arrow sizes. $g^b$ dominates over $\delta g$ for $d=\sqrt{r^2+z^2} \lesssim 3$ kpc, while the converse is true for
$d \gtrsim 3$ kpc. The angle between these vector fields stays small but it is not negligible, except 
in the $z=0$ plane, on the $z$-axis, and far from the matter distribution.

\begin{figure}
\centering
\resizebox{0.43\textwidth}{!}{%
\ \ \ \includegraphics{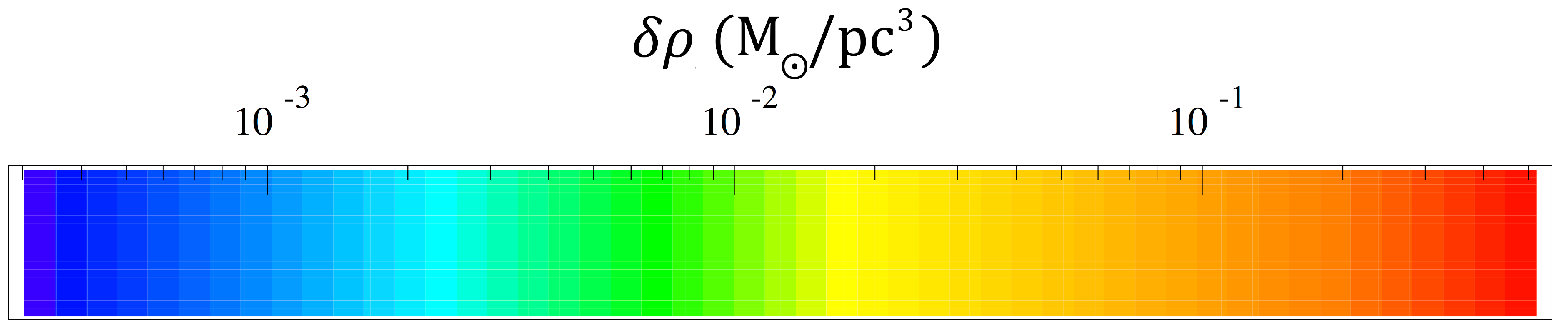}}
\resizebox{0.43\textwidth}{!}{%
\includegraphics{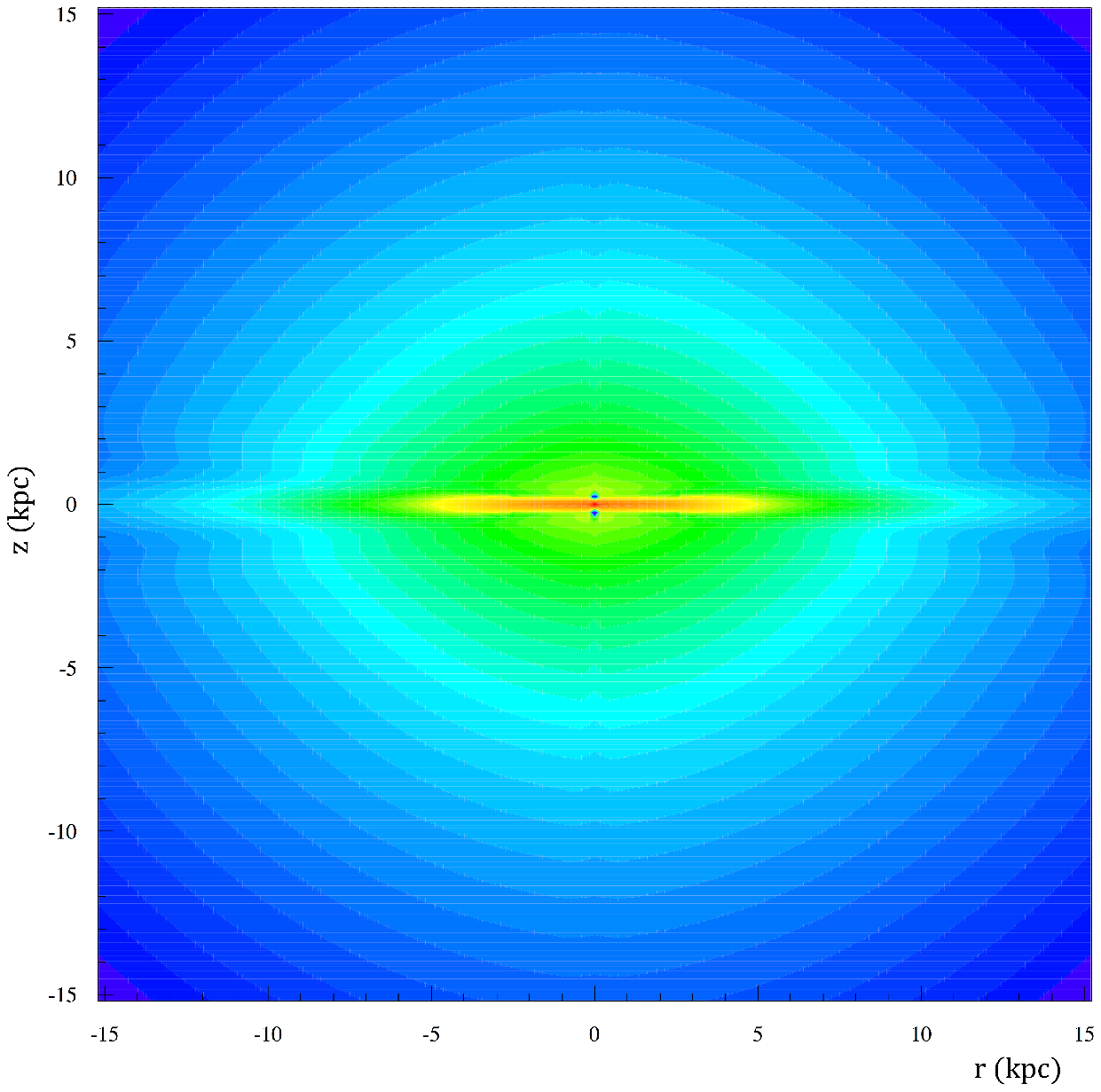}}
\caption {\label{fig:dens2d}{Estimated $\delta\rho$ profile around M33. The horizontal axis
is a cut in the galactic plane, and the vertical axis $z$ is perpendicular to the galactic plane.}}
\end{figure}

Figure \ref{fig:dens2d} shows the $\delta\rho$ distribution around the M33 galactic plane, 
computed from the free-falling velocity field. $\delta\rho$ is maximum inside the matter distribution, and a halo shows up around the disk plane. It is not spherical. Its mass, integrated over a sphere of radius $r=23$ kpc, is $\Delta M_{23}\approx 1.3\times 10^{11} {M}_\odot$. 

\begin{figure}
\centering
\resizebox{0.43\textwidth}{!}{%
\includegraphics{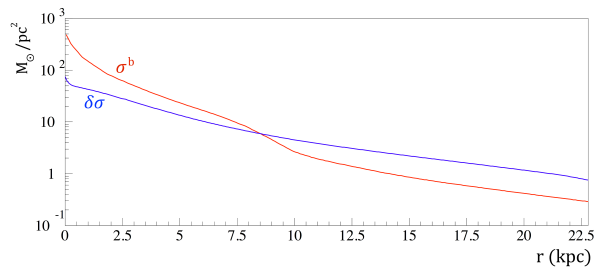}}
\caption {\label{fig:densplan}{$\sigma^b(r)$ (red line) and $\delta \sigma(r)$ (blue line)
are the matter and halo densities integrated over a window $z \in \left[-1,+1\right]$ kpc, where the $z$-direction is transverse to the galactic plane.  $r$ is the galactocentric distance.}}
\end{figure}

So, our model predicts a halo density profile strongly correlated to the real matter one, as does
the MOND paradigm for the ``phantom dark matter" (see for instance Fig. 18b of ref.~\cite{Famaey-2012}). Yet, unlike in the MOND case where $\mathbf{g}$ and $\mathbf{g}^b$ are directly linked by a function, the coupled system summarized in Section \ref{sec:recall}
does not lead to a direct relation between the two accelerations. 

Figure \ref{fig:densplan} compares the real matter and the $\delta\rho$ distributions
integrated between the two planes $z=\pm 1$ kpc, which comprises most of the real matter in the thick
galactic plane. The two surface densities are comparable.
$\delta\sigma$ is peaked at the origin, just like the real matter distribution.

\subsection{THINGS survey}
\label{sec:THINGS}
The H1 Nearby Galaxy Survey~\cite{Walter-2008} acquired precise H1 measurements on a set of 34 Spiral Galaxies. Rotation curves and mass models from a selected set of 19 galaxies have been produced and analysed under the standard dark matter halo
hypothesis~\cite{deBlok-2008}. A selection of the best ones, {\it i.e.} those with the least non-circular star motions, has subsequently been fit under the MOND hypothesis~\cite{Gentile-2010}.

We used the rotation curves and Newtonian velocity components available from~\cite{deBlok-2008} to analyse the set used in~\cite{Gentile-2010} under the hypotheses of our model.
For some galaxies, we removed some data from the fit, following prescriptions given
in~\cite{deBlok-2008} concerning mainly non-circular motions that were detected. 

The parameter $T$ is fixed at the best-fit value acquired from the M33 rotation curve.

As in~\cite{deBlok-2008,Gentile-2010}, we adjust a correction factor to the M/L ratio of each star component,
hereafter denoted ``disk $\Upsilon_c$" for the disk, and ``bulk $\Upsilon_c$" for galaxies
hosting a measured bulk component.

Unlike in the case of M33, where a full mass model extending beyond the measured RC allowed a precise separation of the galactic potential into $\Phi^b$ and $\Phi_{\rm off}$, here the unavailability of such a full mass model does not permit such a precise computation of
the self-potential of the galaxy $\Phi^b(0)$, or to separate $v_{\rm off}$ from
$v_{\rm prox}$ in (\ref{eq:vfree}). So, the other fitted parameter is $v_{\rm prox}(r_{\rm prox})$, where $r_{\rm prox}$ is the distance of the closest point to the galactic centre available in the Newtonian velocity curves of~\cite{deBlok-2008}.
\begin{figure*}[ht]
\centering
\resizebox{.98\textwidth}{!}{%
\includegraphics{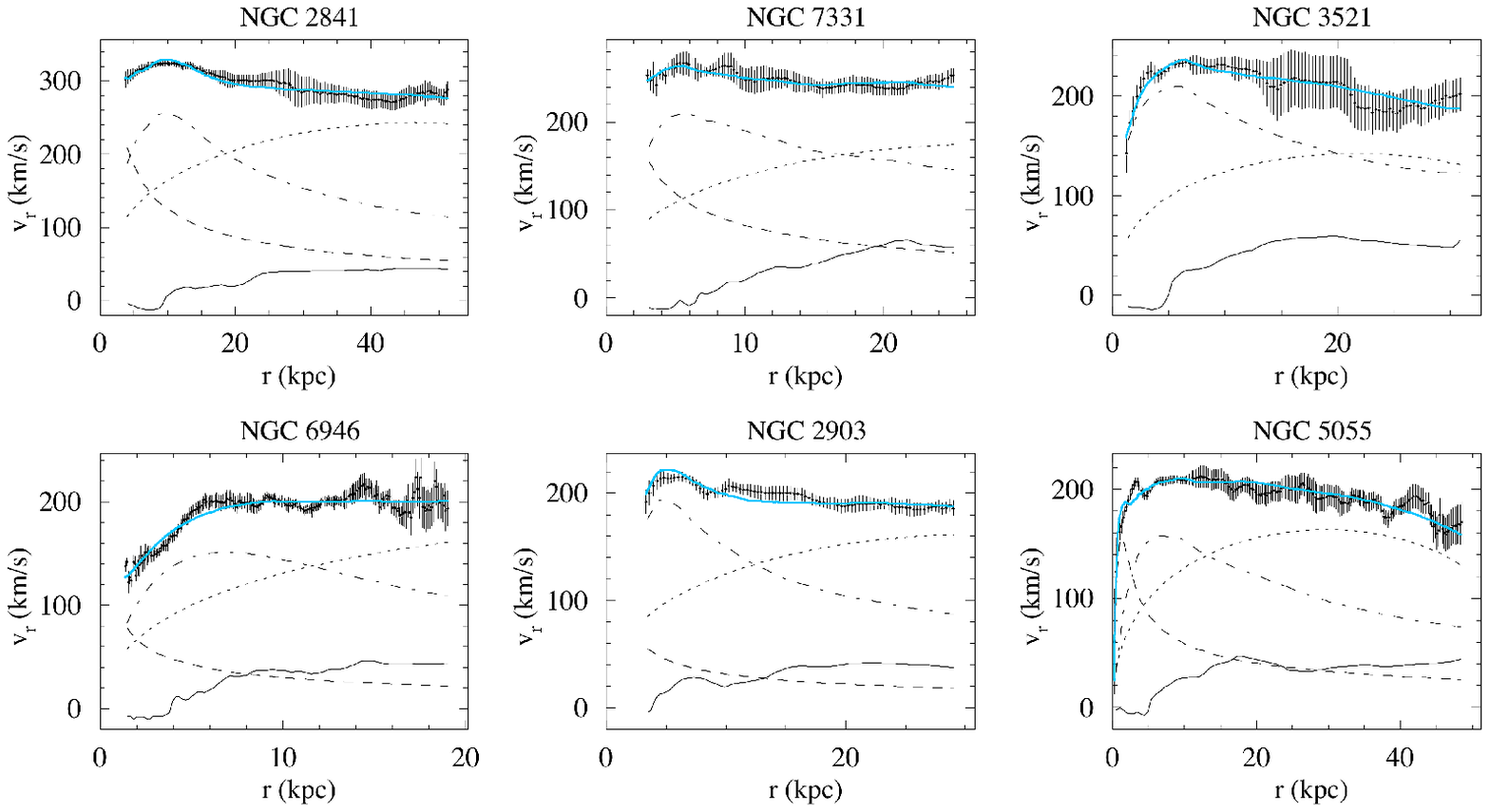}}
\resizebox{.98\textwidth}{!}{%
\includegraphics{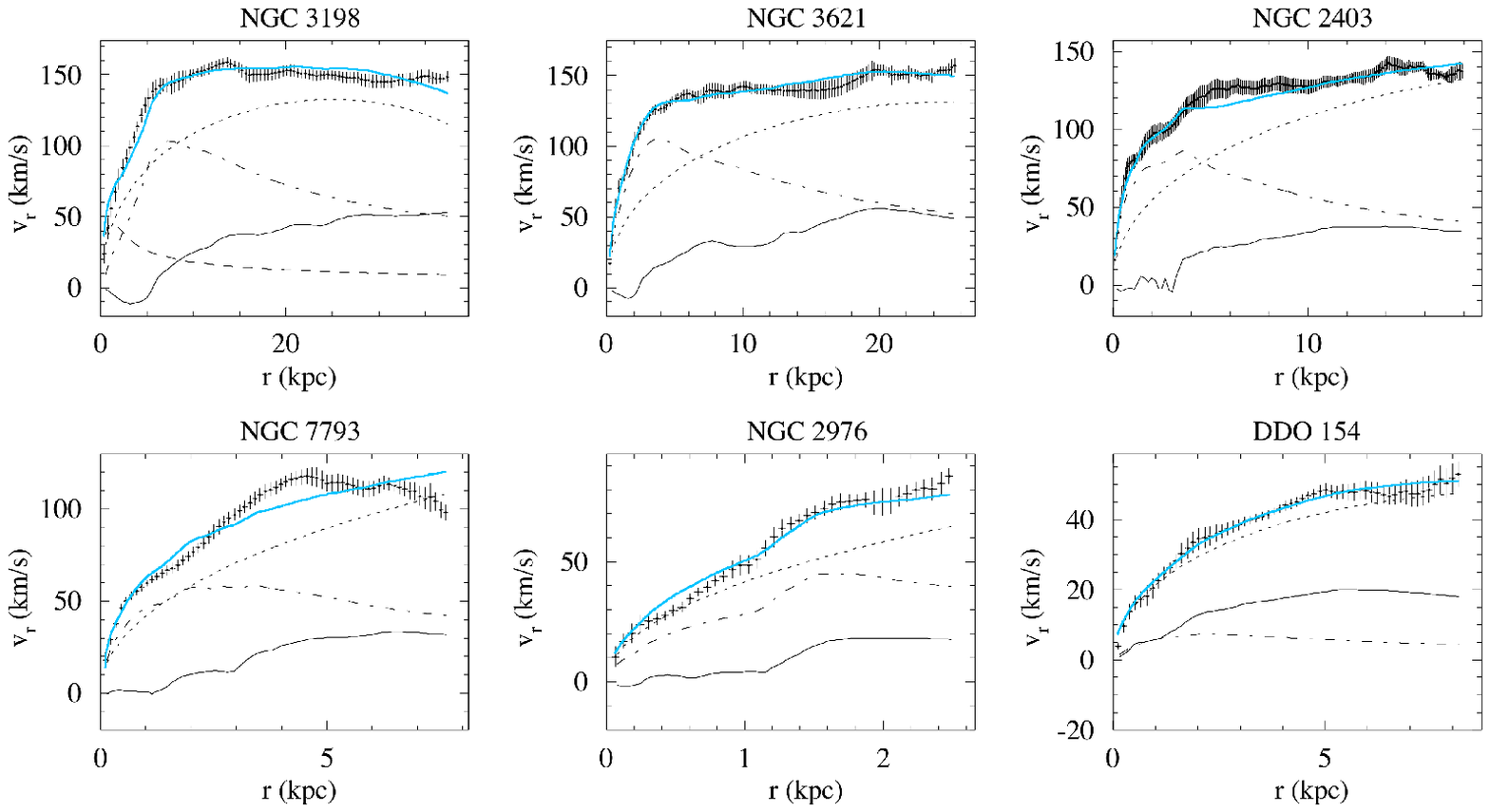}}
\caption {\label{fig:Changalop}{THINGS rotation curve fits, with $T = 6.8 \times 10^{15}$ s fixed at its best-fit value on M33.
Continuous, dash-dotted and dashed lines represent the Newtonian contributions of the
gaseous disk, stellar disk, and bulge, respectively, while the ``falling vacuum" contribution is given by
the dotted line. The blue curves show our best-fit model to the total rotation speed. As in~\cite{deBlok-2008}, negative values of the gas component Newtonian velocity correspond to positive (centrifugal) values of $g^b$ (see text).}}
\end{figure*}

Our results are presented in Figure \ref{fig:Changalop} and Table \ref{tab:1}, which can be
compared with Fig. 5 and Table 2 of Ref.~\cite{Gentile-2010} on the MOND side, and Tables 3 and 4 
of Ref.~\cite{deBlok-2008} (figures relating to distinct galaxies being scattered throughout this detailed publication).

In Figure \ref{fig:Changalop}, negative values of the gas component of the Newtonian rotation speed (shown in solid line) correspond to positive (centrifugal) values of $g^b$. The rotation speed $\sqrt{-r g^b}$ in such cases is formally imaginary; however, we adopt the same convention as in~\cite{deBlok-2008} and plot
$-\sqrt{r g^b}$, which is quadratically subtracted from the other components to get
the full rotation speed.

As stated in~\cite{deBlok-2008,Gentile-2010}, the reduced $\chi^2$ extracted from the fit
should not be considered as an absolute indicator of the fit quality. It it useful as a relative indicator
of the ability to reproduce the data when comparing different approaches. In that sense, our
approach gives similar results to those of previous analyses.
\begin{table}[h]
\caption{Summary of our fit results on the 12 selected THINGS galaxies.
$\chi^2_{\rm r}$ is the reduced chi-squared (see comments in the text).}
\label{tab:1}     

{\centering
\begin{tabular}{|l||c|c|c|l|}
\hline\noalign{}
Galaxy& disk $\Upsilon_c$ & bulge $\Upsilon_c$ 
&$v_{\rm prox}$(km/s) & $\chi^2_{\rm r}$\\
\hline
 NGC 2841 & $1.37\pm .02$ & $1.36 \pm .07$ & $748 \pm3 $ & .45 \\
\hline
 NGC 7331  & $0.61 \pm .03$ &  $0.87 \pm .11$ & $590 \pm 5$  & .28  \\
\hline
 NGC 3521  &  $0.74 \pm .01$  & - & $587 \pm 4 $ & .29\\
\hline
 NGC 6946  & $0.73 \pm .01$ & $0.53 \pm .05$ &$505 \pm 5$ & 1.03 \\
\hline
 NGC 2903 & $3.57 \pm .05$  & 0 &$468\pm 2$  & .85\\
\hline
 NGC 5055  & $0.43 \pm .01$ & $6.6 \pm .3$ & $496 \pm 1 $ & .48 \\
\hline
 NGC 3198  & $0.75 \pm .02$ & $0.25\pm .04$ & $351 \pm 2$ & 1.9 \\
\hline NGC 3621  & $0.81 \pm .01$ & - & $364 \pm 1$ & .64 \\
\hline NGC 2403  & $1.29 \pm .02$ & - & $359 \pm 1$ & 2.3 \\
\hline NGC 7793  & $0.96 \pm .04$ & - & $394 \pm 8$ & 4.2 \\
\hline NGC 2976  & $0.74 \pm .1$ & - & $388 \pm 4$ & 1.8 \\
\hline DDO 154  & $1.3 \pm .9$  & - & $105 \pm 1$ & .54 \\
\hline 
\end{tabular}
}
\end{table}

It is important to note that the fits have been performed with the same $T$ value, despite a great disparity in the characteristics of the sample studied: the baryonic galactic masses vary by nearly 4 orders of magnitude and the asymptotic rotation velocities range from 50 to 300 km/s~\cite{deBlok-2008}. 

\begin{figure*}[ht]
\centering
\resizebox{.8\textwidth}{!}{%
\includegraphics{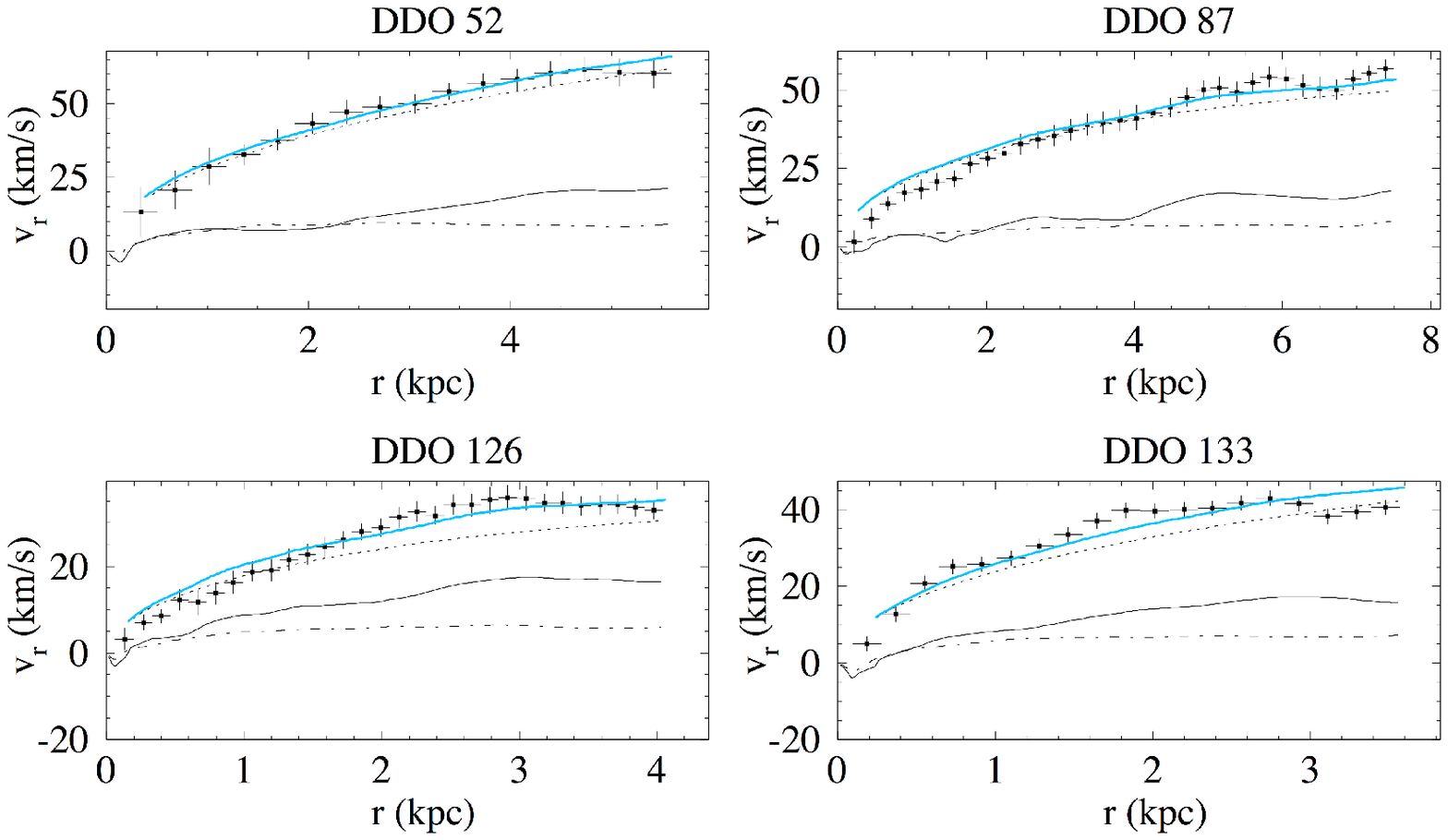}}
\caption {\label{fig:Sand}{Small gaseous galaxies' rotation curve fits with $T = 6.8 \times 10^{15}$ s fixed at its best-fit value on M33.
Continuous and dash-dotted lines represent the Newtonian contributions of the
gaseous disk and stellar disk, respectively, while the ``falling vacuum" contribution is given by
the dotted line. The blue curves show our best-fit model to the total rotation speed. As in~\cite{deBlok-2008}, negative values of the gas component Newtonian velocity correspond to positive (centrifugal) values or $g^b$ (see text).}}
\end{figure*}

\subsection{Small gaseous galaxies}
\label{sec:Gaz}
Dwarf galaxy rotation curves from the LITTLE THINGS survey have been modeled and analysed under the standard dark matter halo hypothesis~\cite{Oh-2015}, and more recently in the MOND framework~\cite{Sanders-2019}.
Their baryonic mass model is dominated by the gaseous component, so their predicted rotation curves are less sensitive to uncertainties on the M/L ratios of the star
components. But they are also less regular, further in many aspects from the ideal
flat, cylindrically symmetric spiral galaxy with no radial movements. In general, they are also less precisely measured and modeled than those of THINGS~\cite{Oh-2015}. We processed them
through our model in the 2D flat cylindrical approximation, although due to their irregular
mass distribution, we cannot consider them as a thorough check of our model (3D mass
models would allow us to compute iteratively the 3D distributions of $\delta\rho$ and better match the
data with model predictions).

For most of these galaxies, the rotation speed does not grow rapidly from the centre,
unlike those of THINGS. Such a behaviour is not surprising since
these object have no centrally peaked star component, while
the halo contribution to the rotation curve does not start to grow from the centre either.
This is the reason these objects became a benchmark for the cusped versus cored halo models 
(see~\cite{Gentile-2007}, for instance).

Some of these galaxies present a central depletion of the gas density that gives rise to an outward-pointing baryonic gravitational field in the vicinity of the galactic centre. This is also the case for some THINGS objects, as can be seen in Figure \ref{fig:Changalop}, but here the gravitational field of the gas disk
is not counterbalanced by a centripetal component produced by a star disk or bulge.
Since, in our model, $\delta \mathbf{g}$ points in the same direction as $\mathbf{g}^b$ in the galactic plane, the extra field enhances this natural centrifugal effect. Unlike $\mathbf{g}^b$ crossing zero when
changing sign, $\delta \mathbf{g}$ presents a discontinuity in direction at the $g^b(r)$ inversion point.

Figure \ref{fig:Sand} and Table \ref{tab:2} show our results for four such objects. The fit takes the surface densities
at face value and adjusts only the $v_{\rm prox}$ parameter.
The $g^b(r)$ inversion point
is poorly known, due to the uncertainties on the M/L star component as well as to the gas
distribution thickness not taken into account in the computation of $g^b$. Nevertheless, our model
matches well the RC data. In the region where $\mathbf{g}$ points outwards, matter is
expelled from the centre and cannot rotate. As in Figure \ref{fig:Changalop}, the celerity plotted on figure \ref{fig:Sand} in this region is computed as $-\sqrt{r g}$ to complement the usual rotation curves where $\sqrt{-r g}$ is represented.

As in the case of THINGS, the error model given in~\cite{Oh-2015} is approximate and the
reduced chi-squared is given only for comparison purposes. It compares well with the NFW
2-parameter dark halo fit given in~\cite{Oh-2015} (chi-squared values are not available in the MOND
paper~\cite{Sanders-2019}, but we agree qualitatively).

{\centering
\begin{table}[h]
\caption{Summary of our fit results on 4 selected LITTLE THINGS galaxies.
$\chi^2_{\rm r}$ is the reduced chi-squared (see comments in the text).}
\label{tab:2}     
\begin{tabular}{|l||c|l|}
\hline
Galaxy& $v_{\rm prox}$(km/s) & $\chi^2_{\rm r}$\\
\hline DDO 52  &  $173 \pm 7$ & .23 \\
\hline DDO 87  &  $104 \pm 3$ & 1.1\\
\hline DDO 126  &  $66 \pm 3 $& 1. \\
\hline DDO 133  &  $124 \pm 4$ & 2.6 \\
\hline 
\end{tabular}
\end{table}
}
%%----------------------------------------------------------------------------------------------------------------------------------------------------------------------------------------------
\section{Conclusion and outlook}
We propose a new paradigm that makes free-falling quantum vacuum responsible for the usual manifestations of dark matter.
The mechanism creates invisible vacuum density inhomogeneities $\delta \rho$ from the response of vacuum fluctuations to the gravitational field. 
In a stationary regime, we assume that the ``vacuum fluid'' corresponding to $\delta\rho$ behaves like a classical conserved medium.
The mechanism is driven by the gravitational potential rather than by the gravitational field itself.  
Surprisingly, the key parameter driving the density inhomogeneities is the divergence of the free-falling flow velocity (and not, for instance, the field itself, as in
a polarisable medium). The $\delta\rho$ field is directly computable from the distribution of ordinary masses $\rho^b(\mathbf{r})$ and an overall additional potential. The coupling between $\delta \rho$ and
the free matter density $\rho^b$ is purely gravitational, and fully constrained by the model.

We are led to introduce a phenomenological characteristic of vacuum, having
the dimensions of time, denoted $T$. Its value, determined by the best fit to the M33 rotation curve data, is found to be
\begin{equation}
T =   (6.8 \pm.6)\times 10^{15}\ {\rm s} =  (220 \pm 20)\, {\rm My}
\label{eq:Time}
\end{equation}

We have presented this approach in the stationary regime in which objects produce a constant gravitational field. We have checked the model on a set of spiral galaxy rotation curves, and it succeeds in
reproducing their characteristics. The halo shapes we predict present similarities with those of ``phantom 
dark matter'' from the MOND paradigm. Both models share nonlinearity as a basic feature.

If our conjecture is correct, it would explain the rotation curves of spiral galaxies while conserving the standard form of Newton's laws, and without calling upon the existence of new stable particles.
We may anticipate that the mechanism can be made responsible for
enhancements of velocity dispersions in elliptical galaxies.  However, exploring this subject will require lengthy 3D simulations.

On the theoretical side, the model needs to be elaborated upon in order to consolidate its foundations, and hopefully it can be
upgraded to a complete microscopic description.

Many other aspects of the dark matter puzzle need to be explored following this paradigm.
We hope this paper is the first of a set in which other aspects of the model are explored, and we hope it will foster independent works on other galaxies
and in other astrophysical and cosmological contexts.

\paragraph{Acknowledgements}

We thank Ma\"el Chantreau, who cho\hyp{}se to dedicate
an L3 internship to this subject and performed outstanding work on the analysis of the THINGS
sample galaxies. 

\paragraph{Declarations}
The authors did not receive support from any organisation for the submitted work.
The authors have no competing interests to declare that are relevant to the content of this article.

\paragraph{Data availability statement}
All data analysed during this study are available in articles cited in reference. No specific
data have been produced for this work.

%\end{Acknowledgements}

\end{document}